\documentclass[12pt,a4paper]{article}
\usepackage{epsfig}
\begin{document}
\title{Flavor changing scalar couplings and $t\gamma(Z)$ production at hadron colliders}
\author{Chong-Xing Yue and Zheng-Jun Zong\\
{\small Department of Physics, Liaoning  Normal University,
Dalian, 116029. P. R. China}
\thanks{E-mail:cxyue@lnnu.edu.cn}}
\date{\today}
\maketitle
\begin{abstract}
\hspace{5mm} We calculate the contributions of the flavor changing
scalar ($FCS$) couplings arised from topcolor-assisted technicolor
($TC2$) models at tree-level to the $t\gamma$ and $tZ$ production
at the Tevatron and $LHC$ experiments. We find that the production
cross sections are very small at the Tevatron with
$\sqrt{s}=1.96TeV$, which is smaller than 5 fb in most of the
parameter space of $TC2$ models. However, the virtual effects of
the $FCS$ couplings on the $t\gamma(Z)$ production can be easily
detected at the $LHC$ with $\sqrt{s}=14TeV$ via the final state
$\gamma l\bar{\nu}b$ ($l^{+}l^{-}l\bar{\nu}b$).
\end {abstract}
\newpage
\noindent{\bf I. Introduction}

The top quark, with a mass of the order of the electroweak
symmetry breaking($EWSB$) scale $m_{t}\approx178.0\pm4.36GeV$[1],
is singled out to play a key role in probing the new physics
beyond the standard model($SM$). The properties of the top quark
could reveal information regarding flavor physics, $EWSB$
mechanism, as well as new physics beyond the $SM$[2]. Hadron
colliders, such as the Tevatron and the $CERN$ $LHC$, can be seen
as top quark factories. One of the primary goals for the Tevatron
and the $LHC$ is to accurately determine the top quark properties,
and to see whether any hint of non-standard physics may be
visible.

The anomalous top quark couplings $tqv$ ($q=c$-or $u$-quarks and
$v=\gamma$, $Z$, or $g$ gauge bosons), which are arise from the
flavor changing ($FC$) interactions, can affect top production and
decay at high energy collider as well as precisely measured
quantities with virtual top contributions. In the $SM$, this type
of couplings vanish at the tree-level but can be generated at the
one-loop level. However, they are suppressed by the $GIM$
mechanism, which can not be detected in the present and near
future high energy experiments[3]. Thus, any signal indicating
this type of couplings is evidence of new physics beyond the $SM$
and will shed more light on flavor physics in the top quark
sector.

Single top quark production is very sensitive to the anomalous top
coupling $tqv$, which can be generated in supersymmetery, topcolor
scenario, and other specific models beyond the $SM$. Thus,
studying the contributions of this type of couplings to single top
production is of special interest. This fact has lead to many
studies involving single top production via the $tqv$ couplings in
lepton colliders[4,5] and hadron colliders[6,7,8].

To completely avoid the problems arising from the elementary Higgs
field in the $SM$, various kinds of dynamical $EWSB$ models have
been proposed, and among which the topcolor scenario is attractive
because it can explain the large top quark mass and provide
possible $EWSB$ mechanism[9]. Almost all of this kind of models
propose that the scale of the gauge groups should be flavor
non-universal. When one writes the non-universal interactions in
the mass eigen-basis, it can induce the tree-level $FC$ couplings.
For example, the top-pions $\pi_{t}^{\pm,0}$ predicted by topcolor
scenario have large Yukawa couplings to the third family quarks
and can induce the tree-level flavor changing scalar ($FCS$)
couplings[10], which have significant contributions to the
anomalous top couplings $tqv$[5]. The aim of this paper is to
calculate the contributions of the $FCS$ coupling
$\pi_{t}^{0}\bar{t}c$ to the processes $gc\rightarrow t\gamma$ and
$gc\rightarrow tZ$ in the framework of topcolor-assisted
technicolor (TC2) models[11], and see whether the effects of the
$FCS$ coupling $\pi_{t}^{0}\bar{t}c$ on $t\gamma$ and $tZ$
production can be detected at the Tevatron and the $LHC$
experiments.

\noindent{\bf II. The calculations of $t\gamma(Z)$ production in
$TC2$ models}

For $TC2$ models[9,11], the underlying interactions, topcolor
interactions, are assumed to be chiral critically strong at the
scale about $1 TeV$ and coupled preferentially to the third
generation, and therefore do not possess $GIM$ mechanism. This is
an essential feature of this kind of models due to the need to
single out top quark for condensate. The non-universal gauge
interactions result in the mass eigen-basis, which can induce the
anomalous top quark couplings $tuv$ and $tcv$. However, the $tuv$
couplings can be neglected because the $FCS$ coupling
$\pi_{t}^{0}\bar{t}u$ is very small[10]. The effective forms of
the anomalous coupling vertices $\Lambda_{tcZ}$,
$\Lambda_{tc\gamma}$, and $\Lambda_{tcg}$ can be writted as[5]:
\begin{equation}
\Lambda^{\mu}_{tcZ}=ie[\gamma^{\mu}(F_{1Z}+F_{2Z}\gamma^{5})+
p_{t}^{\mu}(F_{3Z}+F_{4Z}\gamma^{5})+
p_{c}^{\mu}(F_{5Z}+F_{6Z}\gamma^{5})],
\end{equation}
\begin{equation}
\Lambda^{\mu}_{tc\gamma}=\Lambda^{\mu}_{tcZ}|_{F_{iZ}\rightarrow
F_{i\gamma}},\ \ \ \ \ \
\Lambda^{\mu}_{tcg}=ig_{s}\frac{\lambda^{a}}{2}[\gamma^{\mu}F_{1g}+p_{t}^{\mu}F_{2g}
+p_{c}^{\mu}F_{3g}]
\end{equation}
with
\begin{equation}
F_{i\gamma}=F_{iZ}|_{v_{t}=\frac{2}{3}, a_{t}=0},\ \ \ \ \ \ \
F_{ig}=\frac{3}{2}F_{i\gamma}.
\end{equation}
Where $\lambda^{a}$ is the Gell-Mann matrix. The from factors
$F_{iv}$ are expressed in terms of two - and three - point
standard Feynman integrals[12]. The expressions of $F_{iZ}$ are
given in Ref.[5].

Obviously, the $FCS$ coupling $\pi_{t}^{0}\bar{t}c$ can generate
contributions to the processes $g(p_{g})+c(p_{c})\rightarrow
t(p_{t})+\gamma(p_{\gamma})$ and $g(p_{g})+c(p_{c})\rightarrow
t(p_{t}')+Z(p_{Z})$ via the anomalous top quark couplings
$tc\gamma$, $tcZ$, and $tcg$. The relevant Feynman diagrams are
shown in Fig.1, in which Fig.1(a) and Fig.1(b) come from the
anomalous $tc\gamma$ and $tcZ$ couplings, while Fig.1(c) and
Fig.1(d) come from the anomalous $tcg$ coupling. The renormalized
amplitudes for these processes have similar forms with those of
the process $\gamma p\rightarrow\gamma c\rightarrow t\gamma(Z)$,
which have been given in Ref.[8]. \vspace{-1cm}
\begin{figure}[htb]
\vspace{-2.5cm}
\begin{center}
\epsfig{file=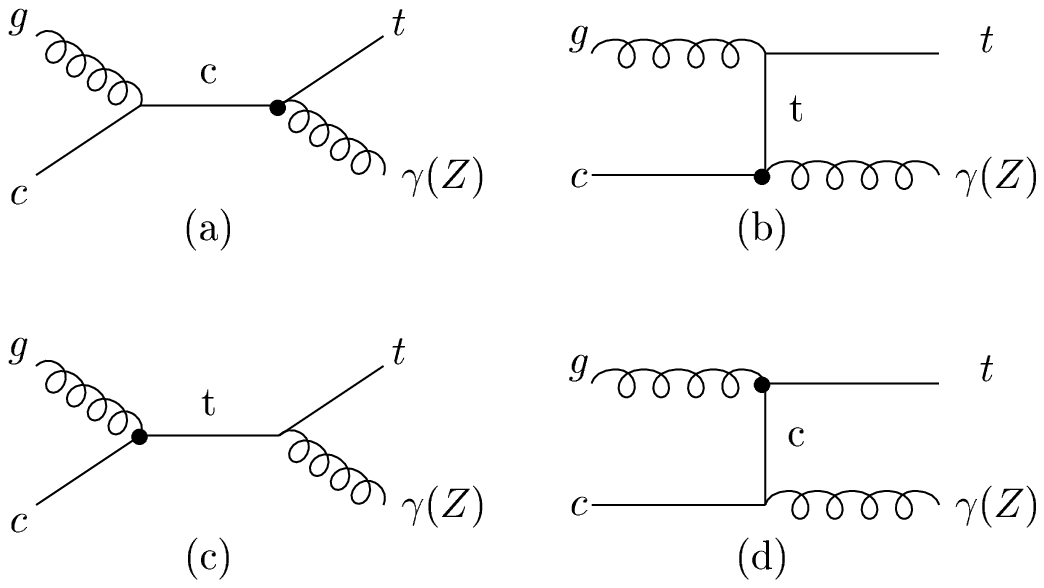,width=550pt,height=800pt} \vspace{-18.5cm}
\hspace{1cm} \vspace{-0.6cm} \caption{Feynman diagrams for
$t\gamma(Z)$ production contributed by the anomalous top
\hspace*{2.0cm}coupling vertices $\Lambda_{tc\gamma},
\Lambda_{tcZ},$ and $\Lambda_{tcg}$.} \label{ee}
\end{center}
\end{figure}

After calculating the partonic cross sections
$\hat{\sigma}_{i}(\hat{s})$ for the subprocesses $gc\rightarrow
t\gamma$ and $gc\rightarrow tZ$, the total cross sections
$\sigma_{i}(s)$ at hadron coliders are obtained by convoluting
$\hat{\sigma}_{i}(\hat{s})$ with the partion distribution
functions $f_{c/p}(x_{1},Q)$ and $f_{g/p}(x_{2},Q)$ of the initial
state particles $c$ and $g$:

\begin{eqnarray}
\sigma_{i}(s)=\int\int
dx_{1}dx_{2}f_{c/p}(x_{1},Q)f_{g/p}(x_{2},Q)\hat{\sigma}_{i}(\hat{s}),
\end{eqnarray}
where $\hat{s}=xs,$ and $x=x_{1}x_{2}$. In our calculation, we
will take the CTEQ5 parton distribution function for
$f_{c/p}(x_{1},Q)$ and $f_{g/p}(x_{2},Q)$ with
$Q^{2}=\hat{s}$[13].

\noindent{\bf III. Numerical results and conclusions}

From above equations, we can see that the cross sections of
$t\gamma$ and $tZ$ production at the Tevatron and the $LHC$ are
dependent on two free parameters $\varepsilon$ and $m_{\pi_{t}}$
of $TC2$ models, except the $SM$ input parameters $\alpha_{e},\
\alpha_{s},\ S_{W},\ m_{Z} $ and $m_{t}$. In $TC2$ models,
topcolor interactions make small contributions to $EWSB$ and give
rise to the main part of the top quark mass,
$(1-\varepsilon)m_{t}$ with $0.01\leq\varepsilon\leq0.1$, a model
dependent free parameter. The limits on the top-pion mass
$m_{\pi_{t}}$ may be obtained via studying its effects on various
observables[9]. It has been shown that $m_{\pi_{t}}$ is allowed to
be in the range of a few hundred $GeV$ depending on the models. As
numerical estimation, we will assume $m_{\pi_{t}}$ and
$\varepsilon$ in the ranges of $200GeV\sim400GeV$ and
$0.01\sim0.1$, respectively.

\begin{figure}[htb]
\centering {
\label{fig:subfig:a}
\includegraphics[width=3.0in]{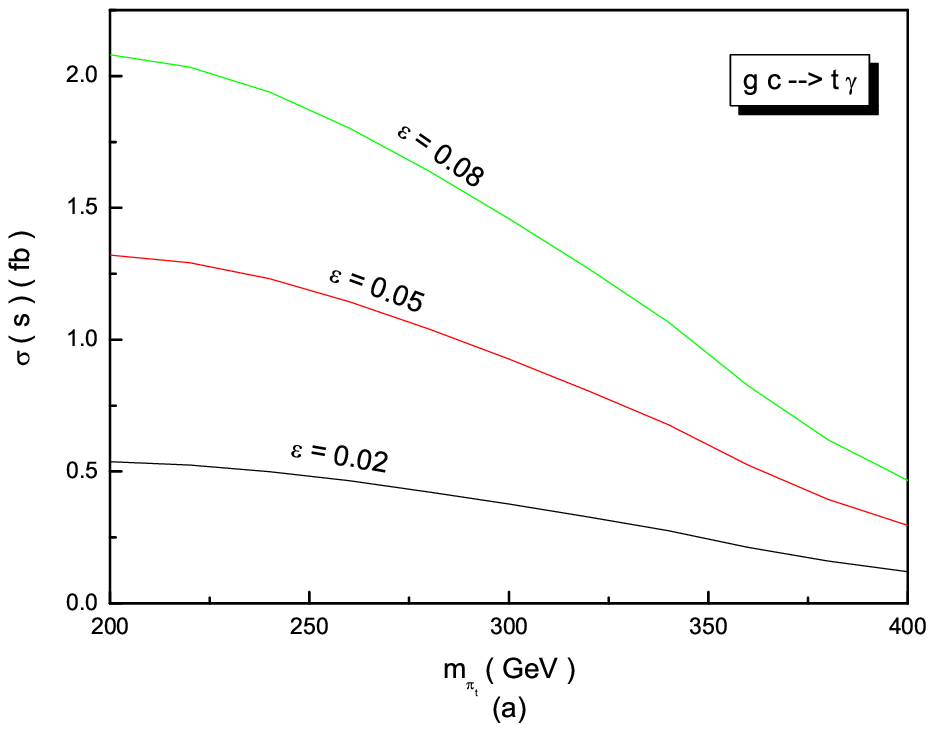}}
\hspace{-0.5cm} {
\label{fig:subfig:b}
\includegraphics[width=3.0in]{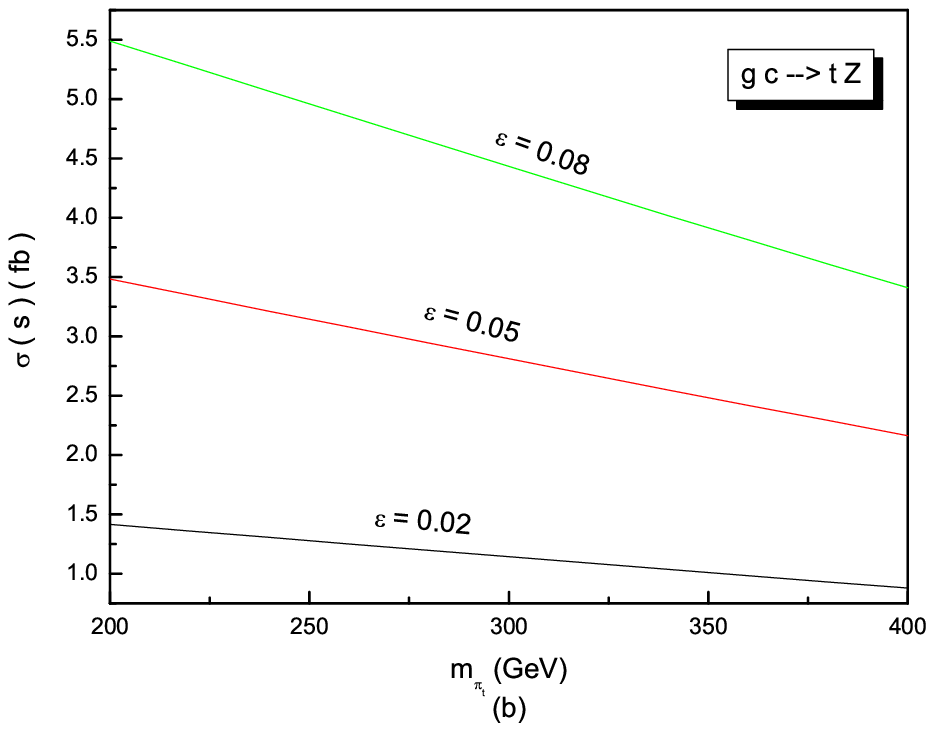}}
\caption{\footnotesize The cross section $\sigma(s)$ of
$t\gamma(Z)$ production as a function of the top-pion mass
$m_{\pi_{t}}$ for three values of the parameter $\varepsilon$ at
the Tevatron with $\sqrt{s}=1.96TeV$ }
\label{fig:figure} 
\end{figure}

\begin{figure}[htb]
\centering {
\label{fig:subfig:a}
\includegraphics[width=2.8in]{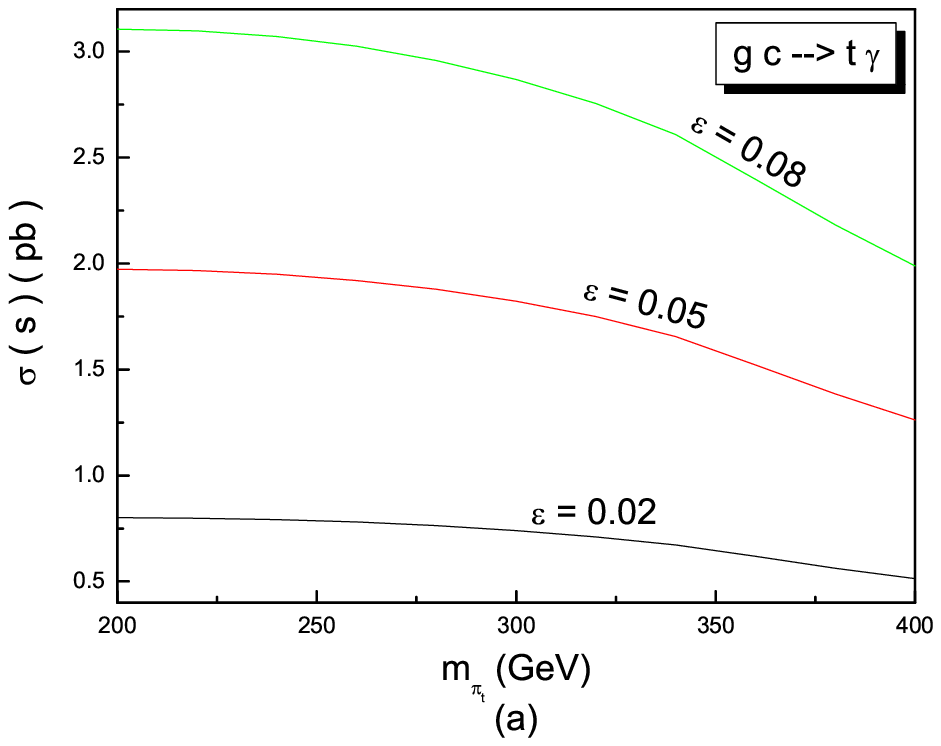}}
\hspace{0.5cm} {
\label{fig:subfig:b}
\includegraphics[width=2.8in]{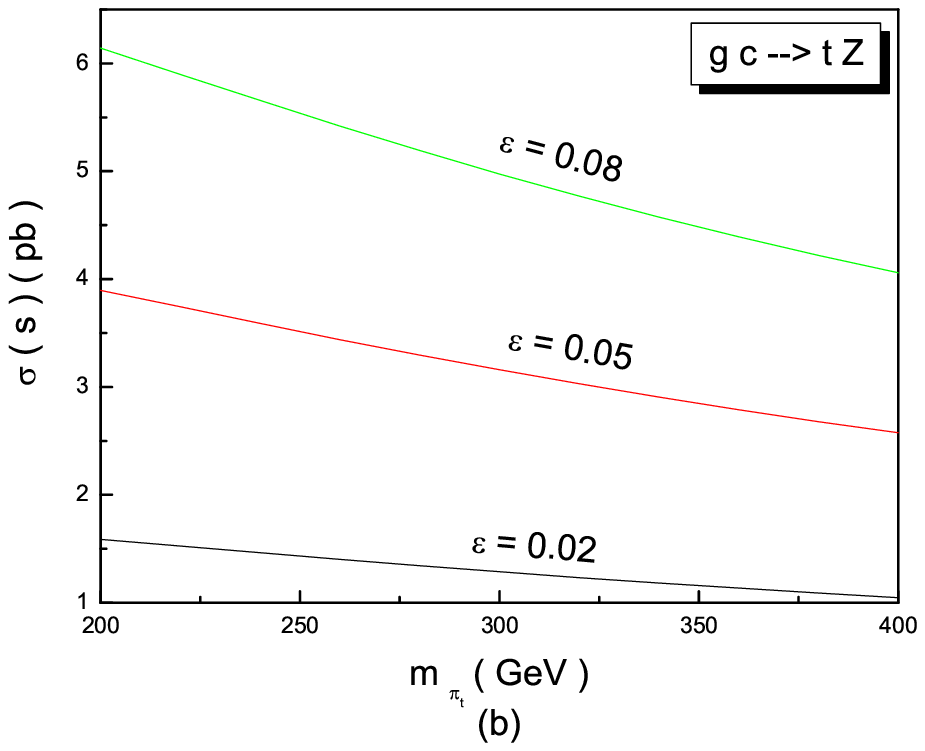}}
\vspace{-0.5cm} \caption{\footnotesize The cross section
$\sigma(s)$ of $t\gamma(Z)$ production as a function of
$m_{\pi_{t}}$ for three values of the parameter $\varepsilon$ at
the $LHC$ with $\sqrt{s}=14TeV$ }
\label{fig:figure} 
\end{figure}

The cross sections $\sigma_{i}(s)$ of $t\gamma(Z)$ production at
the Tevatron with $\sqrt{s}=1.96TeV$ and $LHC$ with
$\sqrt{s}=14TeV$ are plotted as functions of the neutral top-pion
mass $m_{\pi_{t}}$ for three values of the free parameter
$\varepsilon$ in Fig.2 and Fig.3, respectively. From these
figures, we can see that the $t\gamma$ production cross section is
smaller than $tZ$ production cross section at the same collider
experiment. This is because the effective coupling strength of
$\Lambda_{tc\gamma}$ is smaller than that of $\Lambda_{tcZ}$ and
the c. m. energy $\sqrt{s}\gg m_{Z}$. For $200GeV\leq
m_{\pi_{t}}\leq 400GeV$ and $0.02\leq\varepsilon\leq0.08$, the
cross sections of $t\gamma$ and $tZ$ production at the Tevatron
are in the ranges of $1.2\times10^{-4}pb\sim2.1\times10^{-3}pb$
and $8.8\times10^{-4}pb\sim5.4\times10^{-3}pb$, respectively. If
we assume the yearly integrated luminosity
$\pounds_{int}=2fb^{-1}$ for the Tevatron with $\sqrt{s}=1.96TeV$,
then the number of the yearly production events is smaller than 10
in almost of all parameter space of $TC2$ models. Thus, it is very
difficult to detect the effects of the $FCS$ coupling
$\pi_{t}^{0}\bar{t}c$ on the $t\gamma$ and $tZ$ production at the
Tevatron experiments. However, it is not this case for the future
$LHC$ experiment with $\sqrt{s}=14TeV$ and
$\pounds_{int}=100fb^{-1}$. There will be
$3.1\times10^{5}\sim5.1\times10^{4}\ t\gamma$ events and
$6.1\times10^{5}\sim1.0\times10^{5}\ tZ$ events to be generated
per year.

\begin{figure}[htb]
\begin{minipage}[t]{0.5\linewidth}
\centering
\includegraphics[width=8cm]{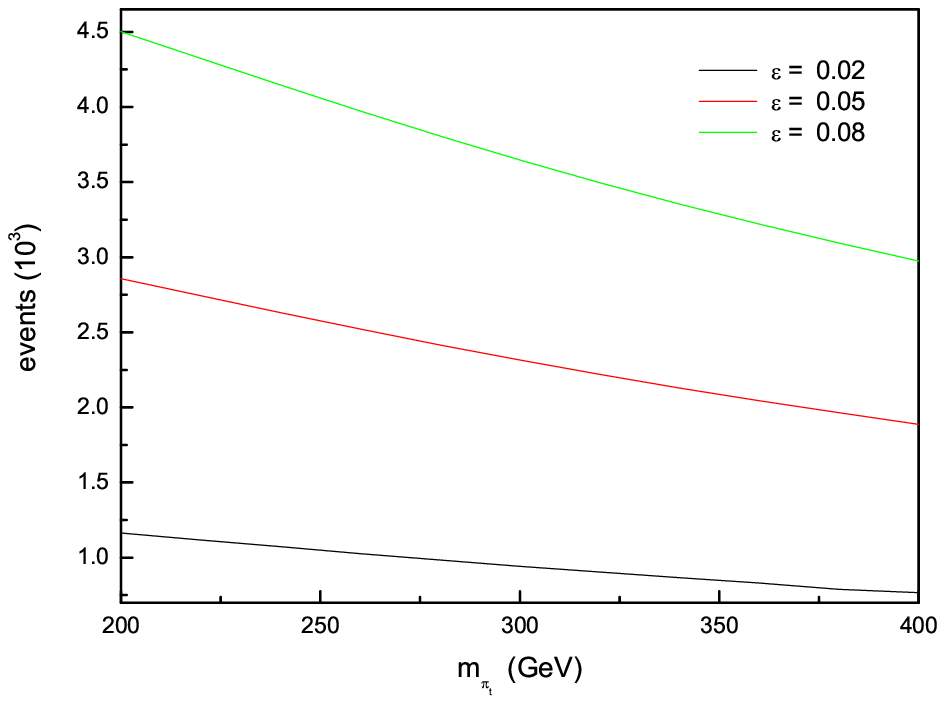}
\caption{\footnotesize The number of the $l^{+}l^{-}l\bar{\nu}b$
events generated at the $LHC$ with $\sqrt{s}=14TeV$ is showed as a
function of $m_{\pi_{t}}$ for three values of the parameter
$\varepsilon$}\label{fig:side:a}
\end{minipage}%
\hspace{0.5cm}
\begin{minipage}[t]{0.5\linewidth}
\centering
\includegraphics[width=8cm]{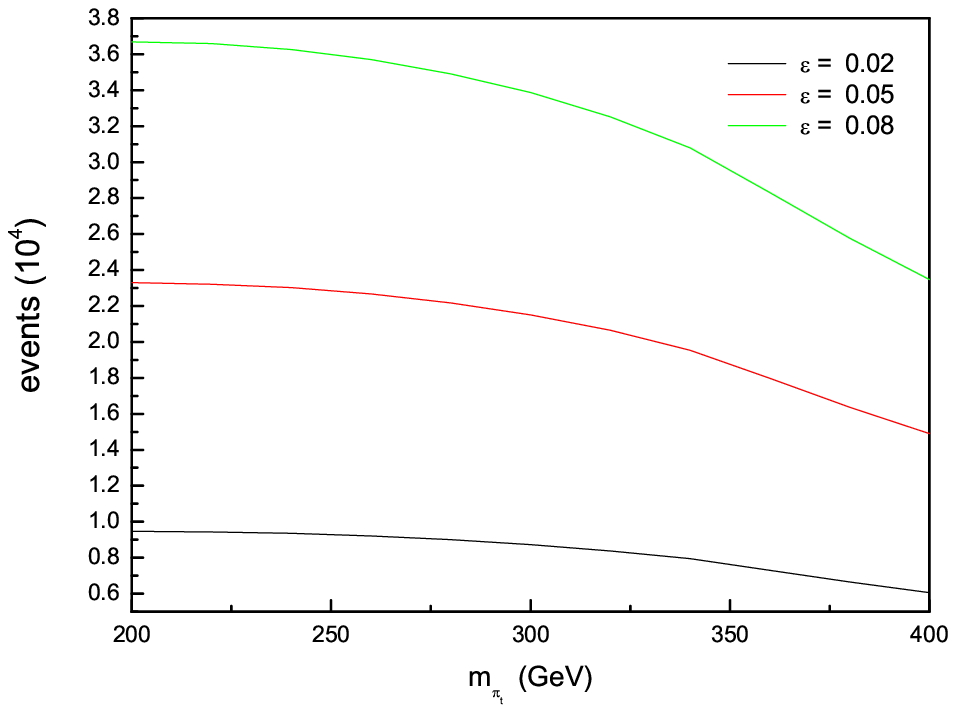}
\caption{\footnotesize The number of the $\gamma l\bar{\nu}b$
events generated at the $LHC$ with $\sqrt{s}=14TeV$ is showed as a
function of $m_{\pi_{t}}$ for three values of the parameter
$\varepsilon$}\label{fig:side:b}
\end{minipage}%
\end{figure}

In general $tZ$ production gives the possible observable five
fermion final states with at least one b quark $ffffb$ including
$\nu\bar{\nu}jjb$ with $Z\rightarrow\nu\bar{\nu}$ and
$W\rightarrow q\bar{q'}$, $jjl\nu b$ with $Z\rightarrow q\bar{q'}$
and $W\rightarrow l\bar{\nu}$, etc. It has been shown that the
final state $\nu\bar{\nu}jjb$ with branching ratio
$B_{r}(tZ\rightarrow \nu\bar{\nu}jjb)\approx 13\%$ is the best
signal event for detection the new physics effects on $tZ$
production at the Tevatron with $\sqrt{s}=1.96TeV$ and
$\pounds_{int}=2fb^{-1}$[14]. Although the final states
$b\bar{b}l\bar{\nu}b$ and $l^{+}l^{-}l\bar{\nu}b$ have smaller
branching ratios $B_{r}(tZ\rightarrow b\bar{b}l\bar{\nu}b)\approx
3.3\%$ and $B_{r}(tZ\rightarrow l^{+}l^{-}l\bar{\nu}b)\approx
1.5\%$, they are the most interesting modes at the $LHC$ with
$\sqrt{s}=14TeV$ and $\pounds_{int}=100fb^{-1}$, due to smaller
backgrounds. In Fig.4, we plot the number of the signal event
$l^{+}l^{-}l\bar{\nu}b$ at the $LHC$ as a function of
$m_{\pi_{t}}$ for three values of the free parameter
$\varepsilon$. In this figure, we have taken the experimental
efficiency $\epsilon$ for detection the final state fermions as
the commonly used reference values: $\epsilon=95\%$ for leptons
and $\epsilon=60\%$ for b quark. One can see from Fig.4 that, in
most of the parameter space of $TC2$ models, there will be several
hundreds and up to thousands observed $l^{+}l^{-}l\bar{\nu}b$
events to be generated at the $LHC$ experiments. Thus, the virtual
effects of the $FCS$ coupling $\pi_{t}^{0}\bar{t}c$ on $tZ$
production should be detected in the future $LHC$ experiments.

Compared with those of $tZ$ production, the final states with a
photon and three fermions of $t\gamma$ production are very simply,
only $\gamma l\bar{\nu}b$ and $\gamma jjb$ depending whether the
$SM$ gauge boson $W$ decays into leptons or hadrons. The leptonic
final state $\gamma l\bar{\nu}b$ has a branching ratio
$B_{r}(t\gamma\rightarrow\gamma l\bar{\nu}b)\approx 21.8\%$ and
the hadronic final state $\gamma jjb$ has a branching ratio
$B_{r}(t\gamma\rightarrow\gamma jjb)\approx 67.8\%$. The $\gamma
l\bar{\nu}b$ final state is the most interesting signal event, due
to its small $\gamma Wj$ background[14]. The number of the
observed $\gamma l\bar{\nu}b$ events is plotted in Fig.5 as a
function of $m_{\pi_{t}}$ for three values of the parameter
$\varepsilon$. One can see from Fig.5 that the number of the
observed $\gamma l\bar{\nu}b$ events are larger than that of the
observed $l^{+}l^{-}l\bar{\nu}b$ events for $tZ$ production in all
of the parameter space. So, the $FCS$ coupling
$\pi_{t}^{0}\bar{t}c$ can be more easy detected via the $t\gamma$
production process than via the $tZ$ production at the $LHC$
experiments.

$TC2$ models also predict the existence of the neutral scalar
top-Higgs $h_{t}^{0}$, which is a $t\bar{t}$ bound and analogous
to the $\sigma$ particle in low energy $QCD$. Similar to the
neutral top-pion $\pi_{t}^{0}$, it can also give rise to the large
effective $tc\gamma$ and $tcZ$ couplings via the $FCS$ coupling
$h_{t}^{0}\bar{t}c$. Our explicit calculation shows that the
effect of the $FCS$ coupling $h_{t}^{0}\bar{t}c$ on the
$t\gamma(Z)$ production is similar to that of the $FCS$ coupling
$\pi_{t}^{0}\bar{t}c$.

In many of the extensions of the $SM$, the $GIM$ mechanism does
not work so well as in the $SM$. The top quark $FC$ interactions
might be predicted in supersymmetery, topcolor scenaric, and other
specific models beyond the $SM$, which can generate significantly
contributions to rare top decays and single top production
processes[2,15]. Thus, these interactions can lead to observable
effects in various high energy colliders[16]. For example, the
large anomalous top couplings $tcv$ generated by the tree-level
$FCS$ couplings $\pi_{t}^{0}\bar{t}c$ or $h_{t}^{0}\bar{t}c$ can
enhance the branching ratios of the rare top decays $t\rightarrow
cv$[17] and the cross sections of single top production at high
energy $e^{+}e^{-}$ collider($LC$)[5] and the $ep$ colliders[8].
In the context of $TC2$ models, the $FCS$ couplings
$\pi_{t}^{0}\bar{t}c$ and $h_{t}^{0}\bar{t}c$ make the cross
section of the process $e^{+}e^{-}\rightarrow \bar{t}c$ in the
range of $0.014fb \sim 0.35fb$ at the $LC$ experiment with
$\sqrt{s}=500GeV$ and that of the process $ep\rightarrow \gamma
c\rightarrow t\gamma(tZ)$ in the range of
$0.14pb\sim1.37pb(0.13pb\sim1.35pb)$ at the $THERA$ collider with
$\sqrt{s}=1000GeV$, which may be detected in these future collider
experiments. Furthermore, Ref.[17] has shown that, in most of the
parameter space of $TC2$ models, there are $B_{r}(t\rightarrow
c\gamma)\sim 1\times 10^{-6}$, $B_{r}(t\rightarrow cZ)\sim 1\times
10^{-4}$ and $B_{r}(t\rightarrow cg)\sim 1\times 10^{-4}$. At the
$LHC$ experiment with $\sqrt{s}=14TeV$ and
$\pounds_{int}=100fb^{-1}$, the production cross section of the
top quark pairs via standard $QCD$ interactions is about $8\times
10^{5}fb$. If we assume that the top quark decays via
$t\rightarrow cv$ with $Z\rightarrow e^{+}e^{-}$ and the antitop
quark decays via $\bar{t}\rightarrow W^{-}\bar{b}$ with
$W^{-}\rightarrow l^{-}\bar{\nu_{l}}$, then, at most, there are
several hundreds observable events to be generated per year. Thus,
the virtual effects of the $FCS$ couplings $\pi_{t}^{0}\bar{t}c$
or $h_{t}^{0}\bar{t}c$ can be more easy detected via the
$t\gamma(Z)$ production process than via the rare top decays
$t\rightarrow cv$ at the $LHC$ experiments.

Topcolor scenario is one of the important candidates for the
mechanism of $EWSB$. A key feature of this kind of models is that
topclor interactions are assumed to couple preferentially to the
third generation and there do not posses $GIM$ mechanism. The
non-universal gauge interactions can induce the $FCS$ couplings
$\pi_{t}^{0}\bar{t}c$ and $h_{t}^{0}\bar{t}c$. If the virtual
effects of the $FCS$ couplings can indeed be detected at the $LHC$
experiments, it will be helpful to test topcolor scenario and
understand $EWSB$ mechanism.

\vspace{0.5cm} \noindent{\bf Acknowledgments}

Z. J. Zong would like to thank Bin Zhang for helpful discussions.
This work was supported in part by the National Natural Science
Foundation of China under the grant No.90203005 and No.10475037
and the Natural Science Foundation of the Liaoning Scientific
Committee(20032101).

\end{document}